\documentclass[twocolumn,aps,prl,amsmath,amssymb,floatfix,superscriptaddress]{revtex4}

\usepackage{dcolumn}
\usepackage{bm}
\usepackage{amssymb}
\usepackage[pdftex]{graphicx}

\begin{document}


\title{Magnetic avalanches in granular ferromagnets: Thermal activated collective behavior}

\author{Gia-Wei Chern}
\affiliation{Department of Physics, University of Virginia, Charlottesville, VA 22904, USA}


\begin{abstract}
We present a numerical study on the thermal activated avalanche dynamics in granular materials composed of ferromagnetic clusters embedded in a non-magnetic matrix. A microscopic dynamical simulation based on the reaction-diffusion process is developed to modeling the magnetization process of such systems. The large-scale simulations presented here explicitly demonstrate inter-granular collective behavior induced by thermal activation of spin tunneling. In particular, we observe an intriguing criticality controlled by the rate of energy dissipation. We show that thermal activated avalanches can be understood in the framework of continuum percolation and the emergent dissipation induced criticality is in the universality class of three-dimensional percolation transition. Implications of these results to the phase-separated states of colossal magnetoresistance materials and other artificial granular magnetic systems are also~discussed.
\end{abstract}

\maketitle

\section*{1. Introduction}


Avalanche phenomena are ubiquitous in Nature~\cite{sethna01}. When systems are driven by external forces, their responses are often in the form of intermittent bursts of activities or avalanches. One well-studied example is the Barkhausen noise in magnetic system which corresponds to avalanches of domain movements. 
As avalanche behaviors often extend over a huge range of sizes, It has  been suggested that avalanche phenomena can be classified into universality classes that are independent of the microscopic details. A canonical example is the random field Ising model (RFIM) which describes a system of interacting Ising spins subject to site-dependent random fields~\cite{sethna93,perkovic95,spasojevic11}. When this Ising system is driven by a magnetic field, it responds in the form of sporadic large-scale spin reconfigurations. These spin avalanches depend strongly on the level of disorder: while the magnetization dynamics is controlled by large system-spanning events at low disorder, small avalanche clusters dominate the magnetic behavior when disorder is increased. Remarkably, the system exhibits a power-law distribution of avalanche sizes at a critical disorder. This nonequilibrium criticality induced by quenched disorder is characterized by a unique set of critical exponents as in equilibrium phase transitions~\cite{perkovic95,spasojevic11}. The RFIM has since been applied to a number of other systems beyond spins.

Conventional wisdom also suggests that the avalanche behaviors are unaffected by thermal fluctuations and are governed by a deterministic dynamics that depends on the static, quenched disorder in the system~\cite{sethna01,sethna93}. In this regard, avalanche is an intrinsic many-body phenomenon in which spin-spin interactions play an indispensable role.
In many systems, thermal effects are indeed negligible as the energy barriers of spin-flips are often too large for thermal activation to be effective.
On the other hand, recent experimental and theoretical studies on magnetic avalanches in molecular magnets have shown that thermal activation is the dominant factor~\cite{suzuki05,minguez05,decelle09,modestov11,subedi13}. In particular, in the process known as the magnetic deflagration, the spin dynamics is dominated by single-site spin tunneling assisted by thermal activation. Consequently the avalanche behavior is mostly controlled by the diffusion and dissipation of thermal energy. Fundamentally, magnetic deflagration can be understood in the framework of  reaction-diffusion process~\cite{minguez06,garanin07,deutsch07} which also describes avalanche phenomena, ranging from phase transition dynamics to pattern formation and propagation of epidemic~waves.

A key component of thermal activated avalanches is the self-sustainability of excess thermal energies that can be used to overcome the energy barriers. When a system is trapped in a meta-stable state, a small perturbation or a quick variation of the external conditions can induce relaxation of a portion of the system through thermal activated tunneling. The energy released from this initial relaxation can be reabsorbed by the rest of the system and used to overcome the barrier again. An avalanche is then ignited when the relaxation process becomes self-sustained. In this scenario, the rate of energy dissipation naturally plays an important factor as thermal activation becomes negligible if a significant amount of the released energy is removed from the system.

Thermal coupling has also been shown to affect the magnetization dynamics of an important class of materials known to exhibit a colossal magentoresistence (CMR), including perovskite manganites and cobaltites~\cite{mahendiran02,hardy03,ghivelder04,fisher04}. 
Temperature-dependent ultrasharp magnetization steps have been observed in several doped Pr$_{0.5}$Ca$_{0.5}$MnO$_3$ manganites.
Moreover, magnetic deflagration induced by surface acoustic waves has been reported in La-based manganites~\cite{macia07}. Recent experiments have also observed unusual sweep-rate dependence of the avalanche behavior in CMR materials~\cite{macia09,vivien}. These observations all point to the important role of thermal activations in the avalanche dynamics of these perovskite oxides. Motivated by these experiments, the goal of this paper is to present a theoretical investigation of thermal activated collective behaviors in CMR-related compounds.

A crucial realization in the study of CMR materials is the key role of heterogeneity in understanding the CMR phenomena~\cite{dagotto01}. For example, doped cobaltites such as La$_{1-x}$Sr$_x$CoO$_3$ have been shown to exhibit a particularly clear form of this inhomogeneity by electron microscopy~\cite{caciuffo99} and nuclear magnetic resonance~\cite{kuhns03}. Relevant to our study here is the small doping regime, where the ground state can be viewed as consisting of nano-scale ferromagnetic metallic clusters embedded in a non-magnetic insulating matrix. Such phase-separated state thus leads to a natural formation of a granular ferromagnet, analogous to the artificial granular systems that exhibits a giant magneto-resistance (GMR) behavior~\cite{sheng73,helman76,xiao92,milner96}. These GMR meta-materials are composed of ferromagentic clusters deposited in a non-magnetic metallic or insulating matrix. Indeed, lightly doped perovskite cobaltites exhibit a hysteretic magnetoresistance behavior with temperature and field dependence similar to the artificial GMR materials~\cite{wu05}.

In the present work we perform microscopic dynamical simulations to investigate the intrinsic effects of thermal activation in the avalanche dynamics of granular ferromagnets. The model of magnetic clusters used in this work is specifically designed such that inter-granular cooperative phenomena are absent in the zero-temperature athermal dynamics. We develop a numerical scheme based on a microscopic implementation of the reaction-diffusion process modified for the granular systems. The large-scale simulations on lattice of $N\sim 10^8$ sites clearly demonstrate thermal activated collective phenomena in the magnetization dynamics of such materials. In addition, we observe an intriguing non-monotonic dependence of the avalanche behaviors on the dissipation coefficient, indicating the existence of emergent criticality controlled by the rate of energy dissipations in granular ferromagnets. A continuum percolation model is then developed to understand the numerical results. In particular, the emergent out-of-equilibrium critical avalanche corresponds to the percolation transition in this picture. 


\section*{2. Model of magnetic clusters}

\begin{figure}
\centering
\includegraphics[width=0.99\columnwidth]{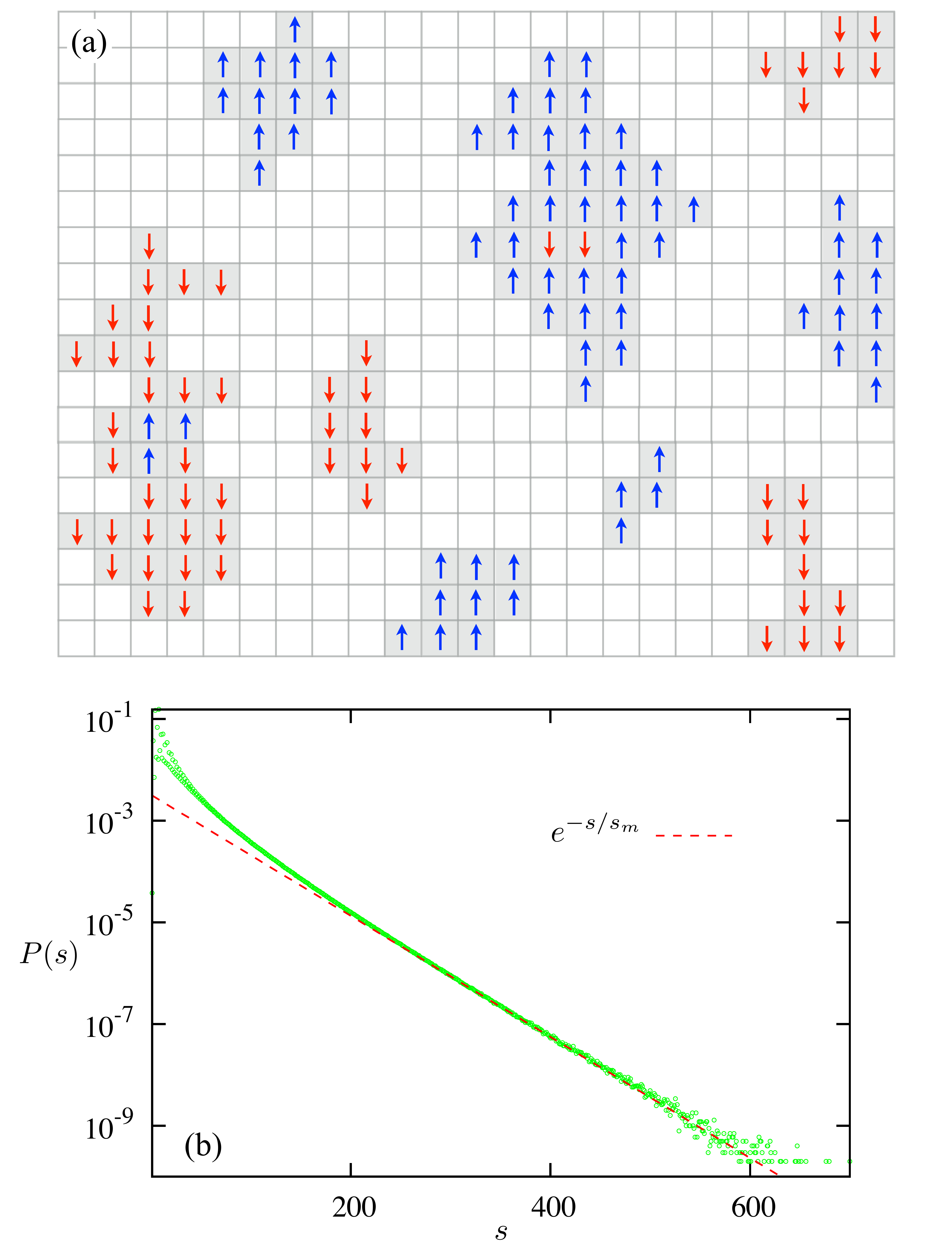}
\caption{(a) Schematic diagram of magnetic clusters embedded in a non-magnetic matrix. The gray cells denote magnetic ions, while white cells represent non-magnetic sites.  The arrows correspond to the Ising variable $\sigma = \pm 1$ of the magnetic ions. (b) Distribution of magnetic clusters generated through a site-percolation process with probability $p$ = 0.05. The distribution at large $s$ follows a exponential decay $P(s) \sim \exp(-s/s_m)$ with $s_m \approx 36$.
\label{fig:hs_cls} }
\end{figure}

The granular ferromagnets considered in this work are composed of magnetic clusters embedded in a non-magnetic matrix.  
In order to perform microscopic simulations, we define the granular model on a simple cubic lattice. A schematic diagram of the granular magnet is shown in Fig.~\ref{fig:hs_cls}(a).
To be more specific, here we will use the case of La$_{1-x}$Sr$_x$CoO$_3$ as the example~\cite{caciuffo99}, although our analysis and conclusion to be discussed in the next section does not depend on details of how the clusters are generated. The parent compound LaCoO$_3$ itself has attracted considerable attention because of an intriguing thermal induced insulator-metal crossover that is accompanied by a spin-state transition of Co ions~\cite{raccah67}. 
 At low temperatures and small fields, the trivalent Co ions are in the non-magnetic low-spin (LS) state with a filled $t_{2g}$ shell. This LS state is energetically close to two other spin-state configurations~\cite{raccah67,korotin96}: an intermediate-spin (IS) state with partially filled $e_g$ and $t_{2g}$ orbitals and spin degree of freedom $S = 1$, and a high-spin (HS) state with active $t_{2g}$ orbitals and $S = 2$. Spin-state crossover can be induced either by enhanced temperatures or the application of a magnetic field.
 
 \bigskip
 
A large magnetoresistance observed in hole doped cobaltites La$_{1-x}$Sr$_x$CoO$_3$  has reinvigorated the research of this perovskite system~\cite{briceno95}. 
The Co ions in the parent compound LaCoO$_3$ occupy the $B$-sites of the perovskite structure and form a cubic lattice. The La atoms sit at the $A$-sites that correspond to the body centers of individual Co cubes.
Substitution of Sr$^{2+}$ for La$^{3+}$  introduces holes into the CoO$_3$ array and locally distorts the lattice such that the neighboring eight Co ions are elevated to the IS state with magnetic moment $S = 1$~\cite{phelan06}. As the holes are confined to the IS clusters, the itinerant electrons interact with the localized spins of IS ions via the Hund's coupling. The metallic IS clusters can thus be thought of as nanoscale ferromagnets stabilized by the double exchange mechanism~\cite{zener51,anderson55}. 

In our simulations, the metallic clusters are generated through a site percolation process~\cite{percolation} as follows. The body-centers of the Co cubic lattice are randomly selected with a probability $p$; a selected body center corresponds to a substituted Sr atom. If the body center of a cube is selected, the surrounding eight Co ions become magnetic. The resultant magnetic clusters are then identified using, e.g. the Hoshen-Kopelman algorithm~\cite{hoshen76}.
Fig.~\ref{fig:hs_cls}(b) shows the probability distribution $P(s)$ of the cluster size averaged over 500 percolation simulations with $p = 0.05$. The distribution shows an exponential decay at large $s$, indicating that the clusters are below the percolation threshold~\cite{percolation}, which is the regime of our main interest as our goal is to study magnetic avalanches assisted by thermal diffusion. For magnetic clusters above the site-percolation transition point, e.g. generated with a large value of $p$, their avalanche dynamics will be dominated by {\em intra}-cluster spin-spin interactions. 
It is worth noting that although we use the case of La$_{1-x}$Sr$_x$CoO$_3$ as the example, our analysis below can be applied to general granular ferromagnets.

For simplicity, we consider Ising spins in this work. An Ising variable $\sigma_i = \pm 1$ is used to denote the spin state of a lattice site that belongs to a magnetic cluster. This is a reasonable approximation for several systems exhibiting easy-axis single-ion anisotropy. Although more realistic models such as Heisenberg spins are possible, the Ising spin approximation allows us to simulate large-scale lattices while capturing the essential physics. 

The most important magnetic property of the clusters is the coercive or threshold field $H_c$. It represents a measure of the external field that is required to demagnetize a fully polarized cluster. While intrinsic single-ion anisotropy is the common source of coercivity, spin-spin interactions also contribute to the coercive field. Spins in magnetic clusters interact with each other through several mechanisms.
First, there is short-range exchange interactions between neighboring spins. For metallic magnetic clusters, the double-exchange mechanism gives rise to a non-local spin interaction that is mediated by the conduction electrons, similar to the well known RKKY interactions. There is also the long-range dipolar interaction for spins of large moment. 
Importantly, significant variation of the coercive field often results from long-range spin-spin interactions, even in the absence of quenched disorder. 
For example, coercivity due to magnetostatic shape anisotropy depends on the geometry of the nano-cluster.

\begin{figure}
\centering
\includegraphics[width=0.99\columnwidth]{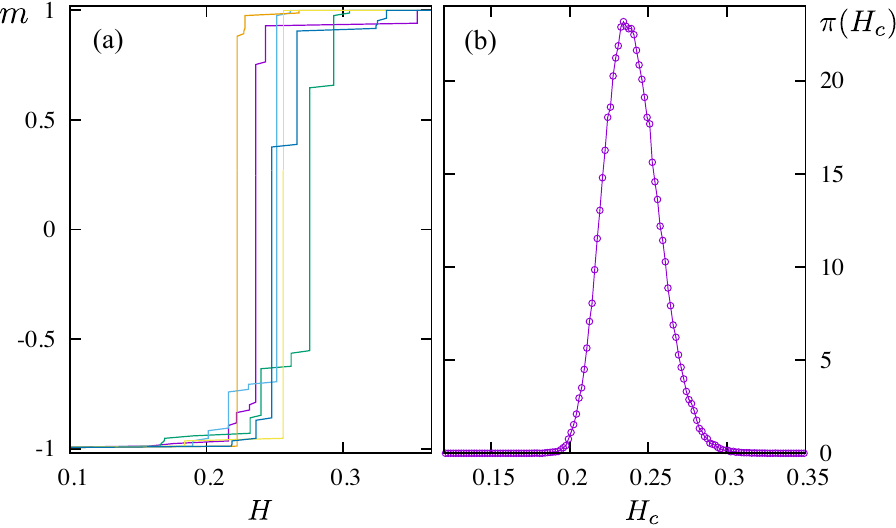}
\caption{(a) Normalized magnetization versus applied field obtained from different random clusters. A coercive field is defined as the field value at which the magnetization shows the largest jump. (b) Distribution of the coercive field obtained from 1000 independent realizations of DE cluster of 180 lattice sites.
\label{fig:de_cls} }
\end{figure}

Metallic magnetic clusters in which electron interact with spins through the double-exchange (DE) mechanism also exhibit a coercive field that depends on the cluster geometry. To demonstrate this property, we simulate the zero temperature magnetization curve of a randomly generated cluster based on the DE mechanism. Our modeling here is particularly relevant for LSCO and other GMR systems containing metallic magnetic clusters. Specifically, we consider a Hamiltonian that corresponds to the strong coupling limit of DE model~\cite{zener51,anderson55}:
\begin{eqnarray}
	\label{eq:de}
	\mathcal{H} = -\frac{t}{2}\sum_{\langle ij \rangle} (1 + \sigma_i \sigma_j)\, (c^\dagger_i c^{\;}_j + {\rm h.c.} ) - H \sum_i \sigma_i.
\end{eqnarray}
Here $\langle ij \rangle$ denotes nearest-neighbor pairs in the cluster, and $c^\dagger_i$ creates an electron with spin parallel to the local Ising moment. The effective hopping constant $t_{ij} = t (1 + \sigma_i \sigma_j)/2$ depends on the relative orientation of spins at the neighboring sites and vanishes when the two spins are antiparallel to each other. The magnetic energy has two contributions: the double-exchange energy and the Zeeman energy corresponding to the two terms in Eq.~(\ref{eq:de}). The DE part is the sum of all filled electron energies $E_{\rm DE}(\{\sigma_i\}) = \sum_{\rm filled} \epsilon_m$, where the eigenenergies $\epsilon_m$ are obtained by directly diagonalizing the tight-binding Hamiltonian for a given spin configuration $\{\sigma_i\}$. For a given external field, the ground state is  obtained by minimizing the total energy, $E_{\rm DE} + E_{\rm Zeeman}$, with respect to Ising variables. Starting from a polarized state with all $\sigma_i = -1$, the magnetization curve of a single cluster is shown in Fig.~\ref{fig:de_cls}(a) for different random realizations. While the demagnetization of such DE clusters is a complex process, the $m(H)$ curves show one or more magnetization jumps at certain field values. We define a coercive field $H_c$ as the value corresponding to the largest magnetization jump. Fig.~\ref{fig:de_cls}(b) shows the probability distribution $\pi(H_c)$ of the coercive field obtained from 1000 independent simulations. A significant variation of the coercive field for DE clusters can be seen from the numerical $\pi(H_c)$. Interesting, the distribution is slightly asymmetric, indicating a non-Gaussian nature of the process.

The above analysis shows that the coercive field of nanoscale clusters in granular magnets depends in general on the size and shape of the cluster. 
For simplicity, we assume in our simulations that each magnetic cluster is characterized by a {\rm single} coercive field $H_c$. The microscopic mechanism for the coercive field varies from one system to another. We also assume that there is an intrinsic disorder in the cluster coercive fields, described by a Gaussian distribution with a mean $\overline{H}_c$ and a variance $\sigma_{H_c}^2$. Again, the microscopic origin of the distribution varies, but most likely involves some long-range interactions.  We expect the general conclusion of our analysis does not depend on details of the coercivity distribution. Since our primary interest in this study is to investigate the intrinsic collective behaviors induced by thermal wave propagation, we will employ the simplest model for the Ising spin dynamics, namely, magnetic behavior of a Ising spin is solely characterized by the coercive field $H_c$ of the cluster to which it belongs. Specifically, this means that at zero temperature an Ising spin will be flipped by an opposite external field only when $|H| > H_c$.

\section*{3. Magnetization dynamics}

Most of the magnetic avalanches, such as Barkhausen and crackling noises, are intrinsically non-equilibrium processes, and are often described by athermal and pure relaxational dynamics. Within this framework, spin inversion takes place whenever the energy of the system can be decreased. The excess energy released during the inversion process is assumed to be quickly transferred out of the magnetic subsystem. The athermal dynamics is often valid at low temperatures and especially when the energy barrier of spin inversion is large enough such that thermal activation can be ignored. Indeed, this approach has been successfully applied to a wide range of magnetic systems. 

However, for systems with small or moderate dissipation, the released energy from spin flip can significantly heat up the lattice locally. When this excess energy spreads over the lattice, further spin inversions might be thermally activated due to the elevated local temperatures. Such scenarios have been discussed recently in the context of magnetic deflagration, in which the spin avalanches are modeled by the highly nonlinear reaction-diffusion process. Mathematically it is described by the following nonlinear partial differential equations~\cite{minguez06,garanin07,deutsch07}:
\begin{eqnarray}
	\label{eq:rd}
	\frac{\partial T}{\partial t} &=& \kappa \nabla^2 T + \frac{\Delta E}{C} \frac{\partial m}{\partial t} - \gamma (T - T_0), \\
	\frac{\partial m}{\partial t} &=& -\Gamma_0 \,e^{-U/T} (m - m_0).
\end{eqnarray}
Here $T$ is the local temperature, $\kappa$ is the thermal diffusivity, $C$ is the heat capacity, $\Delta E$ is the energy released by spin inversion, $\gamma$ is the dissipation coefficient, $T_0$ is the environmental temperature, $m$ is the magnetization density with equilibrium value $m_0$, and $\Gamma_0$ is the intrinsic relaxation rate of the material. The effective relaxation rate is temperature dependent and follows the Arrhenius relation: $\Gamma(T) = \Gamma_0\, \exp(-U/T)$ where $U$ is a barrier energy related to the anisotropy, applied field and other external conditions. It has been shown that a self-sustained propagating spin-reversal front can be generated under appropriate conditions.

Such self-sustained spin reversal could occur within individual nano-clusters of the granular magnet. Here we are interested in the intergranular collective dynamics assisted by thermal energy diffusion. To this end, we introduce a local temperature $T_i$ for each lattice site in addition to the Ising variables (which are defined only for sites belonging to magnetic clusters). Similar to the reaction-diffusion Eqs.~(\ref{eq:rd}) and (3), there are two basic processes in our dynamical simulations, corresponding to the updates of Ising spins $\sigma_i$ and local temperatures $T_i$. 

We first discuss the dynamics of Ising spins. In the presence of an external field $H$, the Ising state that corresponds to spin antiparallel to $H$ becomes meta-stable. The transition from the meta-stable state to the ground-state (with $\sigma_i$ parallel to $H$) is controlled by the energy barrier which is given by $U = H_c - |H|$, where $H_c$ is the coercive field of the cluster. We then assume that the inversion of Ising spin $\sigma_i \parallel -H$ is governed by a Arrhenius dynamics with a transition probability:
\begin{eqnarray}
	\label{eq:w}
	w(\sigma_i \to -\sigma_i) = \min\left\{1, \exp\left(\frac{|H| - H_c}{T_i}\right) \right\}.
\end{eqnarray}
As $T \to 0$, this Metropolis-like dynamics gives rise to a step-like behavior $m \sim \theta(|H| - H_c)$ discussed in the previous section. It is worth noting that despite the seemingly similarity, the Ising dynamics here is not the Metropolis update; once the spin is in the ground state, thermal fluctuations are ignored due to the relatively large energy barrier. Instead, our Ising dynamics is a modified relaxational dynamics which includes thermal activation of the meta-stable state, in much the same spirit as the dynamical equation for $m$ in Eq.~(\ref{eq:rd}).  

\begin{figure*}
\centering
\includegraphics[width=1.98\columnwidth]{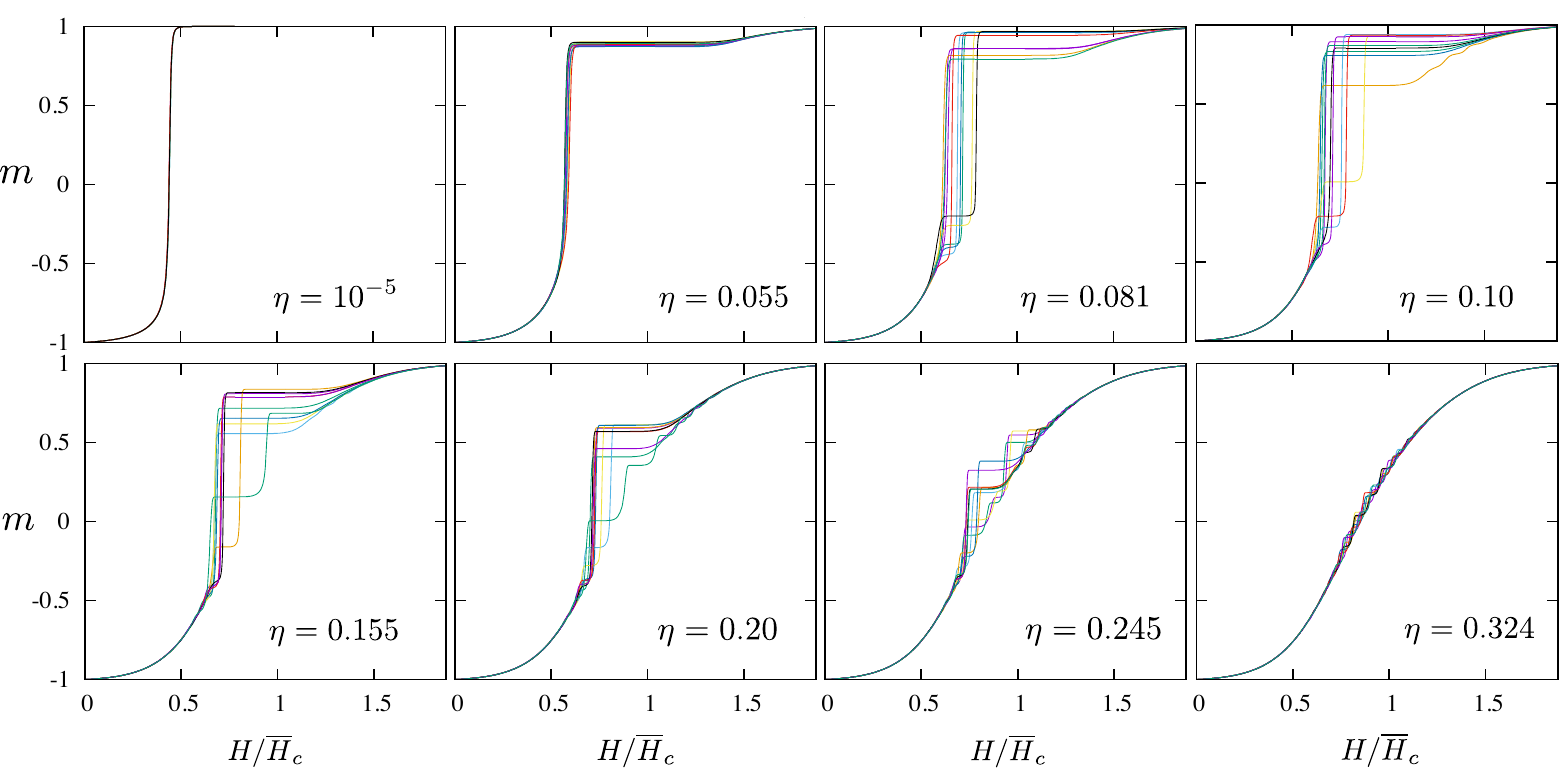}
\caption{Normalized magnetization versus applied field obtained with varying dissipation measure $\eta$. Different curves in each panel are obtained from independent runs with different realizations of the random clusters and their coercive fields. Here $m = \sum_i \sigma_i / M_{\rm max}$ is the normalized magnetization. A large $\eta$ means more energy is dissipated in each field step.
\label{fig:m_h} }
\end{figure*}

Importantly, when the Ising spin $\sigma_i$ is inverted, the released Zeeman energy $2 H$ is deposited at the $i$-th site, which will increase the local temperature by an amount of $2H/C$, where $C$ is the heat capacity.
While the spin updates are only performed on sites belonging to magnetic clusters, the local temperature is updated at every site of the cubic lattice. The update of the temperature variables $T_i$ is governed by the discrete diffusion equation with damping:
\begin{eqnarray}
	\label{eq:dyn_T}
	T_i^{\rm new} = \alpha \, T_i + \mathcal{K} \, \sum_j {\!}' \, (T_j - T_i)  + \Theta\, \delta_i,
\end{eqnarray}
where $\alpha = e^{-\gamma \, \delta t} \approx 1 - \gamma \,\delta t$ is the attenuation coefficient,  $\mathcal{K} = \kappa\, \delta t/\delta x^2$ is the effective diffusivity constant, $\Theta = 2H \delta t / C$ is the temperature increase due to spin flip per time step. We have also set the background temperature $T_0 = 0$. The summation over $j$ only runs over the six nearest neighbors of site-$i$.   In the last term, the notation $\delta_i = 1$  or 0 is used to indicate whether the Ising spin $\sigma_i$ is inverted or not during the time-step.  Consequently, inverted spins act as sources of thermal energy, which then spreads to other lattice sites through the diffusion process.
The dissipation of the thermal energy is controlled by the attenuation coefficient $\alpha$, which plays a crucial role in the thermal activated collective behavior to be described below. 
Since our simulation starts with the initial condition $T_i= 0$, the $\alpha \to 0$ limit corresponds to the zero-temperature athermal dynamics discussed above; all excess energy is quickly removed from the magnetic subsystem.

The dynamical simulation described in Eqs.~(\ref{eq:w}) and (\ref{eq:dyn_T}) thus explicitly takes into account thermal diffusion and dissipation in the magnetization dynamics. We note that within the framework of Monte Carlo simulations, a generalized micro-canonical scheme that incorporates similar thermal effects has been proposed for Ising systems~\cite{deutsch08}. It can be viewed as the Ising equivalent of Langevin dynamics. Our dynamical simulation, on the other hand, should  be viewed as a discretized or lattice implementation of the phenomenological reaction-diffusion dynamics modified for the granular ferromagnets. 

\begin{figure}
\centering
\includegraphics[width=0.92\columnwidth]{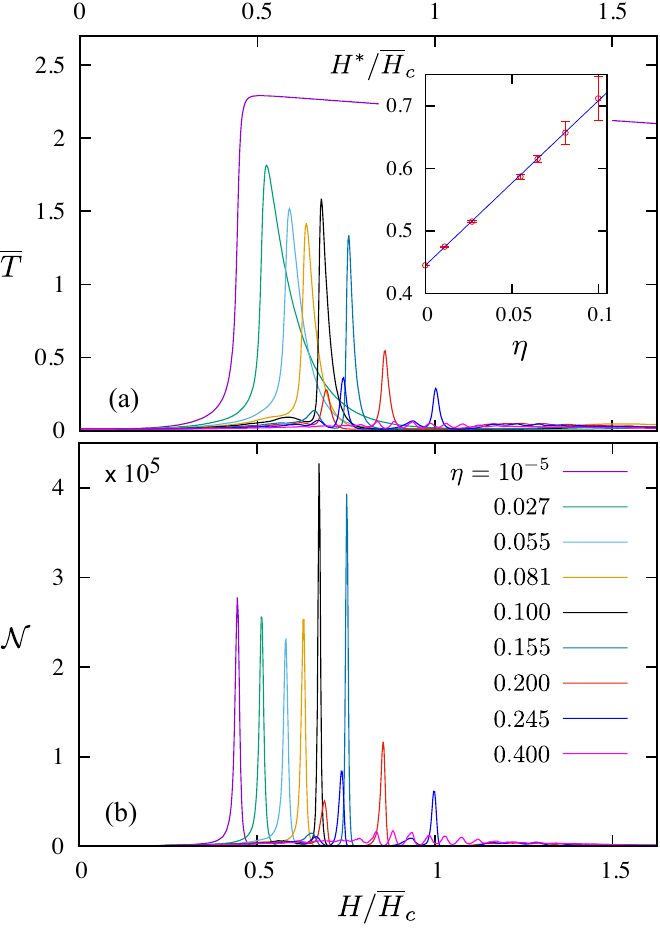}
\caption{(a) Average temperature $\overline{T} = \sum_i T_i / N$ and (b) the number of inverted clusters per field step, $\mathcal{N}$, as functions of applied field for varying degrees of energy dissipation. The peak value of $\mathcal{N}$ shows an intriguing non-monotonic dependence on the dissipation parameter, and the largest avalanche super-cluster appears when $\eta = 0.1$. The inset in (a) shows the  linear dependence of average trigger field $H^*$ on the dissipation parameter $\eta$.
\label{fig:n_cls} }
\end{figure}

\section*{4. Simulation results}

We perform the dynamical simulations on a cubic lattice containing $N = 500^3 \sim 10^8$ sites. At the beginning of each run, all clusters are initially polarized to the negative direction, i.e. $\sigma_i = -1$ for all Ising variables. An external magnetic field $H$ along the positive direction is then slowly ramped up with a constant rate. Numerically, the field-ramping is done by increasing the field with small steps $\Delta H$. This field step corresponds to a time interval $\Delta t = \Delta H / (dH/dt) \equiv \mathbb N \, \delta t$, where $dH/dt$ is the field ramping rate. At each field step, a fixed number $\mathbb N$ of Metropolis  sweeps over spins in all magnetic clusters are carried out. Following each spin sweep, all temperatures are then updated according to Eq.~(\ref{eq:dyn_T}). 
For convenience, we characterize the different simulations by the percentage of energy dissipation at each field step: 
\begin{eqnarray}
\eta = 1 - \exp(-\gamma \Delta t),
\end{eqnarray} 
where the second term is the attenuation of thermal energy due to dissipation $\gamma$. This parameter can be written as $\eta = 1 - \exp(-\gamma \delta t \, \mathbb N) \approx 1 - \alpha^{\mathbb N}$. Experimentally, the rate of field ramp is related to the number of sweeps  used in our microscopic simulations: a larger $\mathbb N$ corresponds to a slower ramping.

In our simulations, we used parameters $\mathbb N = 20$ Monte Carlo and finite-difference updates per field step in our simulations, $\Delta H = 0.002\, \overline{H}_c$, where $\overline{H}_c$ is the average cluster coercive field, the heat capacity $C = 1$, and the thermal conductivity $\mathcal{K} = 0.1$.  Finally, the cluster coercive field is assumed to obey a Gaussian distribution with a standard deviation $\sigma_{H_c} = 0.375 \overline{H}_c$.

Fig.~\ref{fig:m_h} shows examples of magnetization curves obtained from our dynamical simulations with varying degree of dissipation. The curves exhibit step-like features corresponding to magnetic avalanches at small dissipation. The magnetization curves become smoother with increasing $\eta$. In the extreme large dissipation limit $\alpha \to 0$ and $\eta \to 1$, local temperatures $T_i$ remain close to zero throughout the ramping process, indicating the absence of thermal assisted spin tunneling; c.f. Eq.~(\ref{eq:w}) . Spin-flips occur only when $H$ reaches the coercive field of the corresponding cluster. Consequently, inversion of individual magnetic clusters takes place independently of each other and no cooperative behavior is expected. The magnetization simply corresponds to the accumulation of the coercive field distribution, i.e. $m(H)+1 \sim  \int_0^H \pi(H_c) dH_c$. Since here we assume a Gaussian distribution of the random coercive field $H_c$, the resultant magnetization curves resemble a Gauss error function, consistent with the trend shown in Fig.~\ref{fig:m_h}.

\begin{figure*}
\centering
\includegraphics[width=1.99\columnwidth]{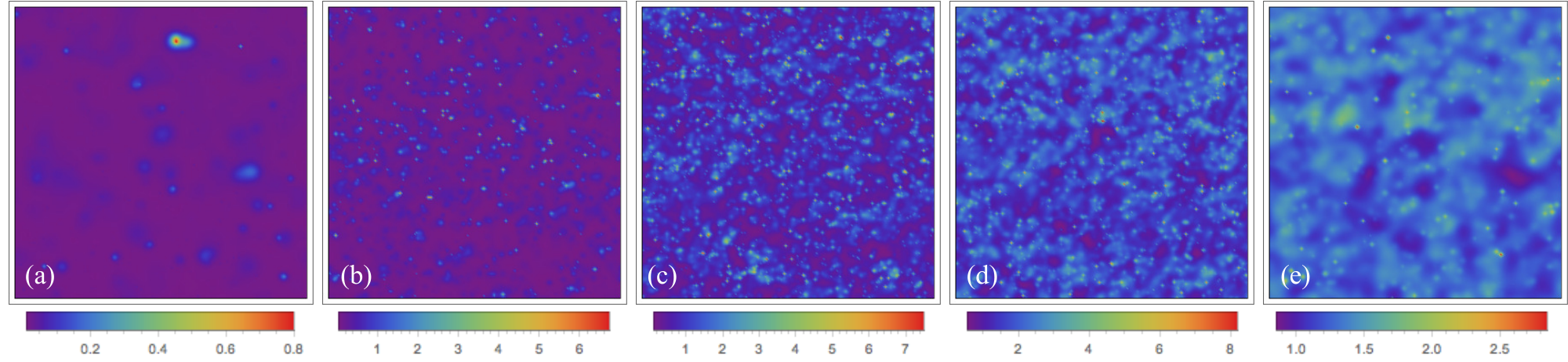}
\caption{Snapshots of the temperature field $T(x,y)$ in a randomly chosen cross section at (a) $H = 0.52 H^*$, (b) $H = 0.989 H^*$, (c) $H = 0.996 H^*$, (d) $H = 1.004 H^*$, and (e) $H = 1.022 H^*$. Here $H^*$ is the trigger field of the avalanche in this simulation with a dissipation parameter $\eta = 0.027$.
\label{fig:snapshots} }
\end{figure*}

It is worth noting that our model is designed in such a way that the system shows no collective behavior in the athermal $\alpha \to 0$ limit. This is in stark contrast to the RFIM and many other systems where the avalanche phenomena are governed by athermal or pure relaxational dynamics. In our model, magnetization jumps that are associated with spin avalanches only occur in the regime of small energy dissipation, indicating that the collective magnetization dynamics is triggered by thermal effects. This thermal activated avalanche phenomenon can be understood as follows. As the field is slowly ramped up, magnetic clusters with smaller coercive field are flipped first. The inversion of these clusters contributes to the smooth shoulder preceding the step-like features, as shown in all panels of Fig.~\ref{fig:m_h}. The diffusion of Zeeman energy released from these leading clusters heats up the system and triggers further spin-flips through thermal activation. Importantly, most of the spins participating in the avalanches belong to subthreshold clusters of which the applied field has yet reached the coercivity. 
This scenario is also consistent with the fact that the observed magnetization jumps mostly occur at field values well below the average coercive field~$\overline{H}_c$.

On the other hand, whether a thermal activated magnetic relaxation can become self-sustained depends crucially on the rate of energy dissipation. As the energy released by the inverted clusters diffuse to neighboring ones, a significant percentage of this energy has to survive the dissipation in order to continue the process. The avalanche behavior of granular magnets thus depends on the efficiency of the matrix serving as a medium to transmit thermal energy. 

To gain more insight into the thermal effects on magnetic avalanches, Fig.~\ref{fig:n_cls}(a) shows the field dependence of the average temperature $\overline{T} = (1/N) \sum_i T_i$.  For the extremely small dissipation case with $\eta = 10^{-5}$, the average temperature increases abruptly at $H \sim 0.5 \overline{H}_c$, and then slowly decays for the rest of the field ramping. All magnetic clusters are inverted during the step-like increase of~$\overline{T}$, again corroborating the thermal activated nature of the magnetic avalanche. For larger dissipations, the average temperature exhibits spikes that correspond to incidents of spin avalanche. Moreover, as the rate of energy dissipation is increased, occurrence of the spikes also shifts toward larger values of the applied field. By averaging the trigger field obtained from tens of independent ramping simulations, we find that the trigger field increases linearly with the dissipation coefficient: $H^* = H^*_0 + A\, \eta$; see the inset of Fig.~\ref{fig:n_cls}(a).

While the magnetic avalanche corresponds to a spike in the average temperature $\overline{T}$, the spatial distribution of local temperatures shows a rather complex pattern. Fig.~\ref{fig:snapshots} shows the temperature profile~$T(x_i, y_i)$ at a randomly chosen cross section of the cubic lattice at different field strengths. At small $H$ compared to the trigger field, there are a few sparsely distributed spots with elevated temperatures. These hot spots correspond to inverted magnetic clusters. The Zeeman energy released from these clusters only heats up the lattice locally. As temperature at these hot spots decays to background $T_0$ with a time constant $1/\gamma$, the excess energy is not enough to trigger further inversion of clusters. On the other hand, as $H$ approaches the trigger field $H^*$, see Fig.~\ref{fig:snapshots}(b)--(d), the number of hot spots increases dramatically. The local temperature rise also increases by an order of magnitude. The significant overlapping of regions with elevated temperatures effectively heats up the whole lattice and triggers a system-size avalanches. It is worth noting that regions with elevated temperatures at $H \sim H^*$ form a complex fractal network which resembles those observed in percolation transitions~\cite{nakayama94}.

Interestingly, detailed examination shows that the largest avalanche event happens at intermediate dissipation parameter $\eta \approx 0.1$, instead of the zero dissipation limit. This is demonstrated in Fig.~\ref{fig:n_cls}(b) which shows the number of inverted clusters per field step, defined as  $\mathcal{N}$, as a function of $H$ for various dissipation parameters. Although the peak value of $\mathcal{N}$ decreases initially as $\eta$ is increased from zero, a non-monotonic behavior is observed and the largest avalanche super-cluster occurs in the case of $\eta = 0.1$.  This unusual dependence on energy dissipation can be attributed to the intricate interplay between the energy dissipation and the random distribution of cluster coercive field. 
In the next section, we discuss this non-monotonic behavior and the existence of a dissipation induced cirticality based on the picture of continuum percolation networks.


\section*{5. Percolation transition}

Before we present the percolation scenario for thermally activated avalanches, we first discuss the linear dependence of the triggering field $H^*$ on the dissipation parameter [inset of Fig.~\ref{fig:n_cls}(a)], which is crucial to the percolation mechanism.  In the continuum limit, the temperature rise due to an inverted cluster can be obtained from solution of the thermal diffusion equation~(\ref{eq:rd}). For simplicity, we assume the cluster is small compared with the thermal diffusion length, hence approximate the source term in Eq.~(\ref{eq:rd}) with a delta-function. The avalanche takes place in the field range $[H^*-\varpi_H, H^* + \varpi_H]$, where the width $\varpi_H$ is rather small; it is approximately a few tens of the field steps from Fig.~\ref{fig:n_cls}. The duration of the avalanche is then given by $t^* = \varpi_H / (dH/dt) $. The temperature field during this time interval is then 
\begin{eqnarray}
	\label{eq:Tr}
	T(\mathbf r) \approx \frac{2H^*}{C} \exp\left(\frac{-r^2}{2\kappa t^* }\right) \,\exp\left(-\gamma \, t^* \right).
\end{eqnarray}
We have assumed the cluster is located at $\mathbf r = 0$.  As in standard diffusion phenomena, we can define a length scale characterizing the range of thermal diffusion:
\begin{eqnarray}
	\label{eq:ell_kp}
	\ell_\kappa = \sqrt{2 \kappa t^*} = \sqrt{2 \mathcal{K} \, t^* / \delta t },
\end{eqnarray}
Here we have used the definition $\mathcal{K} = \kappa \delta t / \delta x^2$, and set the spatial discretization $\delta x = 1$. The avalanche duration corresponds to roughly $t^* / \delta t \approx 1000$ Monte Carlo sweeps in our simulation. 

\begin{figure}
\centering
\includegraphics[width=0.9\columnwidth]{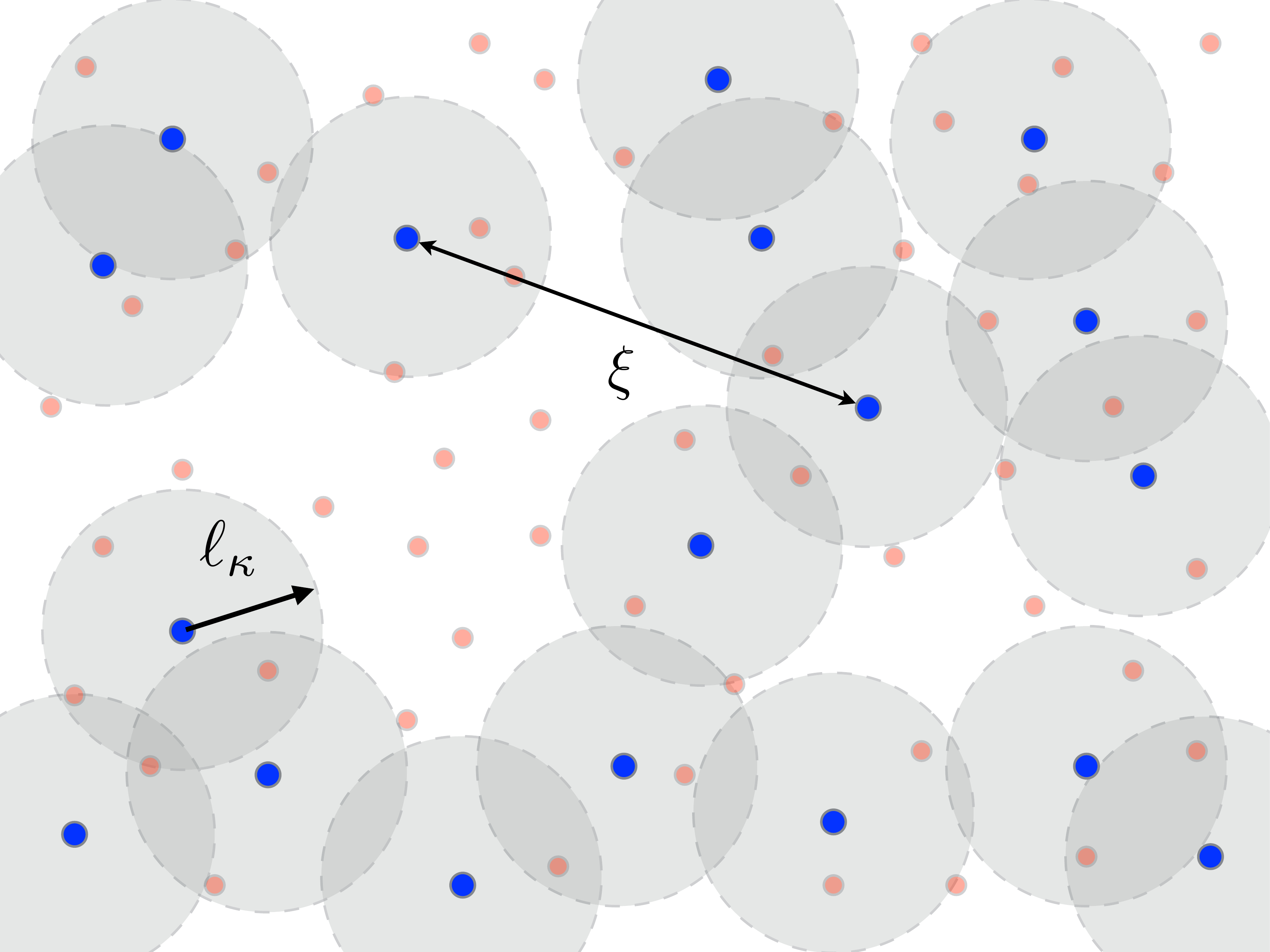}
\caption{Schematic diagram showing the percolation scenario of thermal activated avalanches. The blue circles denote clusters which are inverted by a field $H^*$, while the red circles represent magnetic clusters yet to be inverted with a larger coercive field $H_c > H^*$. The inverted clusters act as heat sources. The diffusion of released thermal energy is characterized by a length scale $\ell_\kappa$. The average distance between inverted clusters is $\xi$, which is a function of $H^*$.
\label{fig:percolation} }
\end{figure}

The physical meaning of this length scale can be understood as follows. For a given inverted magnetic cluster, the released Zeeman energy effectively heats up a sphere of radius $\ell_\kappa$ during a avalanche; see Fig.~\ref{fig:percolation}. Based on Eq.~(\ref{eq:Tr}), the characteristic temperature of the sphere is estimated to be $T^* \sim \mathcal{Q} H^* e^{-\gamma t^* }$, which the constant $\mathcal{Q}$ accounts for the geometric details of the cluster distribution.
The elevated temperature within this sphere should be sufficient to thermally induce inversion of other clusters with larger coercive field in order to self-sustain the avalanche. A simple criterion, a necessary but not a sufficient condition, for the survival of an avalanche is that the temperature rise must overcome the average energy barrier $\overline{U}_b = \overline{H}_c - H^*$ for the majority of clusters; see Eq.~(\ref{eq:w}). The condition $T^* \sim \overline{U}_b$ then provides an estimate of the trigger field:
\begin{eqnarray}
	H^* \sim \overline{H}_c / (1 + \mathcal{Q}\, e^{-\gamma t^*}).
\end{eqnarray}
Using the identity $\exp(-\gamma t^*)  = (1 - \eta)^{t^*/\Delta t} \approx 1 - (t^*/\Delta t) \eta$, we obtain a linear dependence of the trigger field on the dissipation parameter
\begin{eqnarray}
	H^* \sim \frac{\overline{H}_c}{1 + \mathcal{Q}} + \frac{\mathcal{Q}}{(1 + \mathcal{Q})^2} \frac{t^*}{\Delta t} \, \eta.
\end{eqnarray} 
for small $\eta$, consistent with the numerical result shown in the inset of Fig.~\ref{fig:n_cls}(a).
Physically, this dependence can be understood as follows. In the thermal activation scenario, the avalanche is initiated by some seed clusters whose coercive field $H_c \lesssim H^*$. On the other hand, the majority of clusters have a larger coercivity since we assume a normal distribution of the coercive fields. How efficient these seed clusters can thermally induce inversions of other sub-threshold clusters depends on the average temperature rise $T^*$ with the thermal diffusion length $\ell_\kappa$.  This temperature $T^*$ is required to overcome the difference between the coercivity of seed clusters (which is roughly $H^*$) and that of the majority of clusters (which is represented by $ \overline{H}_c$). Consequently, a larger dissipation gives rise to a smaller $T^*$, leading to a trigger field $H^*$ that is closer to the average coercivity. 

The efficiency of the thermal activation for sustaining the avalanche also depends on the number of seed clusters. Here we can employ the continuum percolation to understand this effect. Continuum percolation model has found numerous applications in disordered systems ranging from porous media, composite materials, polymers and colloids~\cite{nakayama94}. The standard model consists of a system of spatially uncorrelated, equal-sized spheres, whose centers are randomly distributed in a three dimensional bulk. The spheres form clusters when they contact or overlap with neighboring spheres. In the case of porous media, these clusters of spheres form the porous paths through the system. As the density of spheres increases, some of the clusters start to span the whole system, signaling the percolation transition.

 In our case, the percolating spheres can be thought of as the spherical region which centers at the inverted cluster and has a radius of thermal diffusion length $\ell_\kappa$; see Fig.~\ref{fig:percolation}. Since the elevated temperature within the sphere can induce further inversions of clusters with larger coercive fields, a system-size percolating super-cluster (cluster of magnetic clusters in our case) corresponds to a magnetic avalanche in which an extensive number of clusters are flipped through thermal tunneling. Whether a system-size super-cluster  can be formed thus depends on the thermal diffusion length $\ell_\kappa$ as well as the density of the seed clusters. 


For an avalanche taking place at $ H^*$, the density of seed clusters $n_c(H_c)$ whose coercive field $H_c \sim H^*$ follows a normal distribution. This introduces a new length scale 
\begin{eqnarray}
	\xi \sim \frac{1}{[n_c(H^*)]^{1/3}} = \xi_0 \,\exp\left[\frac{(H^* - \overline{H}_c)^2}{6 \sigma_H^2}\right],
\end{eqnarray}
which is basically the average distance between seed clusters. Here $\xi_0$ is related to the average density of all magnetic clusters. The percolation threshold in the case of spheres corresponds to a critical ratio $(\ell_\kappa / \xi)_c \approx 0.4338$~\cite{elam84,lorenz01}. The radius $\ell_\kappa$ as given by Eq.~(\ref{eq:ell_kp}) only depends on the thermal conductivity and the duration of avalanches which is not affected much by dissipation. On the other hand, as discussed above, a larger dissipation gives rise to a higher trigger field $H^*$, which in turn results in a higher density of seed clusters and shorter mean distance $\xi$. Consequently, as dissipation rate is increased, the system will undergo a percolation transition when the mean distance between seed clusters reaches the critical value $\xi_c$ determined by the special $H^* = H^*(\eta_c)$, signaling a magnetic avalanche transition.

\begin{figure}
\centering
\includegraphics[width=0.95\columnwidth]{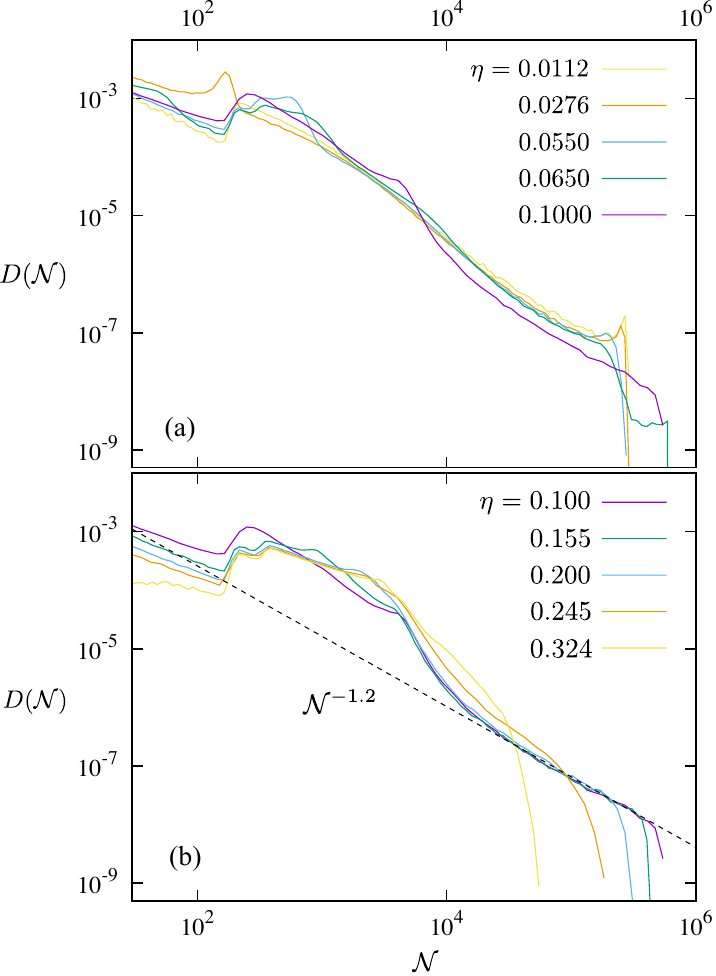}
\caption{Distribution of the avalanche super-clusters in (a) the super-critical regime characterized by weak energy dissipation, and (b) the sub-critical regime corresponding to strong energy dissipation.
\label{fig:dist_cls} }
\end{figure}

For percolating systems, a correlation length can be defined as the characteristic linear size of clusters. Similar to equilibrium phase transitions controlled by temperature, this correlation length diverges at the percolation transition~\cite{nakayama94}. 
The absence of a characteristic length scale at the critical point $\eta_c$ also implies a power-law distribution of super-cluster sizes, which is indeed observed in our numerical simulations; see Fig.~\ref{fig:dist_cls}. For small dissipation, the distribution $D(\mathcal{N})$ shown in Fig.~\ref{fig:dist_cls}(a) exhibits a power-law dependence followed by a spike at largest $\mathcal{N}$, resembling the distribution in the super-critical regime of the RFIM~\cite{perkovic95,spasojevic11}. In conventional supercritical avalanches, the spike occurs at the same largest value bounded by the system size. On the other hand, the spike shown in Fig.~\ref{fig:dist_cls}(a) shifts toward smaller values of $\mathcal{N}$ while its peak also diminishes as $\eta$ increases. At the critical value $\eta_c \approx 0.1$, the distribution shows a power-law dependence all the way to the largest sizes.
Above the critical $\eta_c$, the largest avalanches are cut off by a characteristic size that becomes progressively smaller with increasing dissipation; see Fig.~\ref{fig:dist_cls}(b). The distribution in this regime is again similar to the sub-critical phase of the RFIM~\cite{perkovic95,spasojevic11}. 

In percolation models, the power-law distribution of cluster size at the critical point is characterized the so-called Fisher's exponent $\tau$:
\begin{eqnarray}
	D(\mathcal{N}) \sim \mathcal{N}^{\, 1 - \tau}
\end{eqnarray}
Our simulations find a critical distribution $D(\mathcal{N}) \sim \mathcal{N}^{-1.2}$, as shown by the dashed line in Fig.~\ref{fig:dist_cls}(b), indicating an exponent $\tau \approx 2.2$. This value is  consistent with the Fisher exponent obtained from both continuum and lattice percolation transitions in three dimensions~\cite{lorenz01,lorenz98}. This agreement further corroborates the continuum percolation picture for the thermal activated avalanches.  It is worth noting that unlike the disorder-induced criticality in random-field Ising model~\cite{perkovic95,spasojevic11}, the appearance of a critical distribution in our system is controlled by the rate of energy dissipation.

\section*{6. Summary and discussion}

To summarize, we have conducted large-scale simulations of avalanche dynamics in granular ferromagnets, emphasizing the role of thermal activated inter-granular cooperative behaviors. We have implemented a microscopic simulation scheme based on the reaction-diffusion dynamics. In this framework, the reaction corresponds to local spin flipping, from which the released magnetic energy can diffuse to other lattice sites and trigger further spin inversions through thermal activation. Our simulations have shown that magnetization jumps in granular materials correspond to simultaneous inversions of a huge number of magnetic clusters triggered by such a self-sustained thermal activated relaxation. It is worth noting that magnetic interactions among clusters such as the long-range dipolar force are not included in our modeling of the granular system, the inter-cluster collective behaviors are completely of thermal origin. The matrix serves as a medium that transmits the thermal energy between clusters. We have also systematically studied the effects of dissipation on the avalanche dynamics and uncovered an intriguing criticality controlled by the rate of energy dissipation. The nature of this nonequilibrium critical behavior can be understood within the framework of continuum percolation theory. Essentially, regions with elevated temperatures around the seed clusters overlap with each other and form a percolating network. The emergent criticality thus corresponds to the percolation transition of these high-temperature regions.

Our work was motivated by recent experiments showing unusual sweep-rate dependence of magnetic avalanches in perovskite manganites and cobaltites, especially a non-monotonic dependence of the avalanche dynamics on the sweep rate in LaSrCoO$_3$~\cite{vivien}. Both materials are known to exhibit the novel CMR effect. The phase-separated regime of these compounds, which is crucial to their observed novel properties, can be viewed as granular systems consisting of nano-scale metallic ferromagnets embedded in a non-magnetic matrix. Our analysis can also be applied to granaular GMR meta-materials. Since ferromagnetism in clusters of CMR-related compounds arises from the double-exchange (DE) effect, we have studied the relaxational magnetization dynamics of such DE clusters. 
It is worth noting that magnetic avalanches in large double-exchange clusters is by itself an interesting and important topic. In particular, for metallic granular magnets in the vicinity of or above the percolation threshold, such as highly doped cobaltites, the magnetic clusters merge to form extended regions coexisting with the non-metallic regions. We expect interesting physics to result from the interplay between thermal activation and double-exchange effects.

\section*{Acknowledgment} The author thanks Shalinee Chikara and Vivien Zapf for sharing the unpublished experimental results and for numerous insightful discussions.


\end{document}